\documentclass[aps,prl,twocolumn,superscriptaddress,longbibliography]{revtex4-1}
\usepackage{color}
\usepackage[usenames,dvipsnames]{xcolor}
\usepackage[normalem]{ulem}

\usepackage{graphicx}
\usepackage{epsfig}
\usepackage{dcolumn}
\usepackage{bm}
\usepackage{mathrsfs}
\usepackage{multirow}
\usepackage[all]{xy}
\usepackage{pbox}
\usepackage{amssymb}

\usepackage{color}
\definecolor{myblue}{RGB}{65,105,225}
\definecolor{mygreen}{RGB}{34,139,34}
\definecolor{myorange}{RGB}{255,69,0}
\usepackage[colorlinks,linkcolor=myorange,urlcolor=myblue,citecolor=myblue]{hyperref}

\def\(({\left(}
\def\)){\right)}
\def\[[{\left[}
\def\]]{\right]}

\newcommand{\beq}{\begin{equation}}
\newcommand{\eeq}{\end{equation}}

\newcommand{\ben}{\begin{eqnarray}}
\newcommand{\een}{\end{eqnarray}}

\newcommand{\lan}{\left\langle}
\newcommand{\ran}{\right\rangle}

\newcommand{\la}{\langle}
\newcommand{\ra}{\rangle}
\newcommand{\be}{\begin{equation}}
\newcommand{\ee}{\end{equation}}

\newcommand{\ket}[1]{\left| #1 \ran}
\newcommand{\bra}[1]{\lan #1 \right|}
\newcommand{\bracket}[2]{\lan #1 \right| \!\left. #2 \ran}

\newcommand{\jumpdir}{q}

\begin{document}  

\title{Building continuous time crystals from rare events}

\author{R. Hurtado-Guti\'errez}
\email[]{rhurtado@onsager.ugr.es}
\affiliation{Departamento de Electromagnetismo y F\'{\i}sica de la Materia, Universidad de Granada, Granada 18071, Spain}
\affiliation{Institute Carlos I for Theoretical and Computational Physics, Universidad de Granada, Granada 18071, Spain}

\author{F. Carollo}
\email[]{federico.carollo@uni-tuebingen.de}
\affiliation{Institut für Theoretische Physik, Universität Tübingen, Auf der Morgenstelle 14, 72076 Tübingen, Germany}

\author{C. P\'erez-Espigares}
\email[]{carlosperez@ugr.es}
\affiliation{Departamento de Electromagnetismo y F\'{\i}sica de la Materia, Universidad de Granada, Granada 18071, Spain}
\affiliation{Institute Carlos I for Theoretical and Computational Physics, Universidad de Granada, Granada 18071, Spain}

\author{P.I. Hurtado}
\email[]{phurtado@onsager.ugr.es}
\affiliation{Departamento de Electromagnetismo y F\'{\i}sica de la Materia, Universidad de Granada, Granada 18071, Spain}
\affiliation{Institute Carlos I for Theoretical and Computational Physics, Universidad de Granada, Granada 18071, Spain}

\date{\today}

\begin{abstract}
Symmetry-breaking dynamical phase transitions (DPTs) abound in the fluctuations of nonequilibrium systems. Here we show that the spectral features of a particular class of DPTs exhibit the fingerprints of the recently discovered time-crystal phase of matter. Using Doob's transform as a tool, we provide a mechanism to build classical time-crystal generators from the rare event statistics of some driven diffusive systems. An analysis of the Doob's smart field in terms of the order parameter of the transition then leads to the time-crystal lattice gas (tcLG), a model of driven fluid subject to an external \emph{packing field} which presents a clear-cut steady-state phase transition to a time-crystalline phase characterized by a matter density wave which breaks continuous time-translation symmetry and displays rigidity and long-range spatio-temporal order, as required for a time crystal. A hydrodynamic analysis of the tcLG transition uncovers striking similarities, but also key differences, with the Kuramoto synchronization transition. Possible experimental realizations of the tcLG  in colloidal fluids are also discussed.
\end{abstract}

\maketitle 

\textbf{\emph{Introduction.}}-- 
Most symmetries in nature can be spontaneously broken (gauge symmetries, rotational invariance, discrete symmetries, etc.), with the system ground state showing fewer symmetries than the associated action. A good example is the spatial-translation symmetry, which breaks spontaneously giving rise to new phases of matter characterized by crystalline order, accompanied by a number of distinct physical features such as rigidity, long-range order or Bragg peaks \cite{chaikin00a}. Time-translation symmetry, on the other hand, seemed to be special and fundamentally unbreakable. This changed in 2012, when Wilczek and Shapere proposed the concept of time crystals \cite{wilczek12a, shapere12a}, i.e. systems whose ground state spontaneously breaks time-translation symmetry and thus exhibits enduring periodic motion. This concept, though natural, has stirred a vivid debate among physicists, leading to some clear-cut conclusions \cite{zakrzewski12a, moessner17a, richerme17a, yao18a, sacha18a}. Several no-go theorems have been proven that forbid time-crystalline order in equilibrium systems under rather general conditions \cite{bruno13a, nozieres13a, watanabe15a}, though time crystals are still possible out of equilibrium. In particular, periodically-driven (Floquet) systems have been shown to display spontaneous breaking of \emph{discrete} time-translation symmetry via subharmonic entrainment \cite{khemani16a, keyserlingk16a, else16a, gambetta19a, yao17a}. These so-called discrete time crystals, recently observed in the lab \cite{yao17a,zhang17a,choi17a}, are robust against environmental dissipation \cite{nakatsugawa17a, lazarides17a, gong18a, tucker18a, osullivan18a, lazarides19a} and have also classical counterparts \cite{yao18b,gambetta19b}. In any case, the possibility of spontaneous breaking of \emph{continuous} time-translation symmetry remains puzzling (see however \cite{iemini18a,medenjak20a,buca19b,kozin19a}).

\begin{figure}
\includegraphics[width=8.5cm]{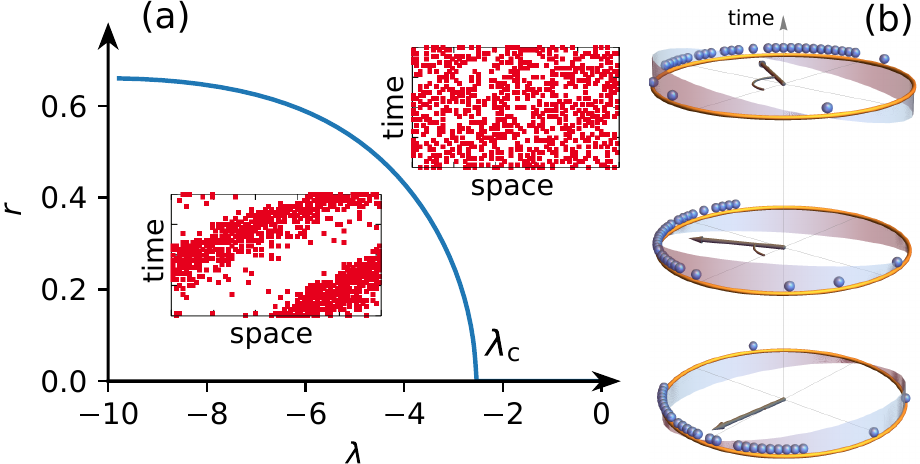}
\vspace{-0.2cm}
    \caption{(a) Packing order parameter $r(\lambda)$ for the DPT in $1d$ WASEP as a function of the biasing field $\lambda$. Inset: spacetime trajectories for current fluctuations above (top) and below (bottom) the critical point. Note the density wave in the latter case. (b) Time-crystal lattice gas with a packing field (shaded curve) which pushes particles lagging behind the center of mass while restraining those moving ahead, a mechanism that leads to a a rotating condensate. The arrow locates the condensate center of mass, with a magnitude $\propto r_C$.
}
\label{fig1}
\end{figure}

Here we propose an alternative route to search for time-crystalline order in classical settings, based on the recent observation of spontaneous symmetry breaking in the dynamical fluctuations of many-body systems \cite{bertini05a, bodineau05a, harris05a, bertini06a, bodineau07a, lecomte07c, garrahan07a, garrahan09a, hurtado11a, ates12a, perez-espigares13a, harris13a, vaikuntanathan14a, mey14a, jack15a, baek15a, tsobgni16a, harris17a, lazarescu17a, brandner17a, karevski17a, carollo17a, baek17a, tizon-escamilla17b, shpielberg17a, baek18a, shpielberg18a, perez-espigares18b, chleboun18a, klymko18a, whitelam18a, vroylandt19a}. Such fluctuations or rare events concern time-integrated observables and are highly unlikely to occur, since their probability decays exponentially with time, thus following a large deviation principle \cite{touchette09a}. However, when these fluctuations come about, they may lead to dynamical phase transitions (DPTs), which manifest as drastic changes in the trajectories of the system and have been recently found in many contexts \cite{bertini05a, bertini06a, derrida07a, hurtado14a, lazarescu15a, shpielberg18a,perez-espigares19a}. In particular, second-order DPTs are associated with the emergence of symmetry-broken structures \cite{bodineau05a, hurtado11a, perez-espigares13a, karevski17a, harris05a, harris13a, chleboun18a, jack15a, carollo17a, carollo18a}.
%In particular, second-order DPTs are associated with the emergence of symmetry-broken structures, and have been found in plenty of systems \cite{bodineau05a, hurtado11a, perez-espigares13a, karevski17a, harris05a, harris13a, chleboun18a, jack15a, carollo17a, carollo18a}.
This is the case of a paradigmatic classical model of particle transport: the weakly asymmetric simple exclusion process (WASEP) in $1d$ \cite{spitzer70a, derrida98a, derrida98b, golinelli06a, perez-espigares13a, hurtado14a, tizon-escamilla17b}. The periodic WASEP is a driven diffusive system that, in order to sustain a time-integrated current fluctuation well below its average, develops a jammed density wave or rotating condensate to hinder particle transport and thus facilitate the fluctuation \cite{bodineau05a,perez-espigares13a}.
This is displayed in the insets to Fig.~\ref{fig1}.a \cite{perez-espigares13a}, where a rotating condensate arises for a subcritical biasing field $\lambda<\lambda_c$, which drives the system well below its average stationary current---corresponding to $\lambda=0$.
%This DPT is captured by a packing order parameter $r$ (which is explained below) that we introduce in order to measure particles' coherent motion, see Fig.~\ref{fig1}.a.
This DPT is captured by a packing order parameter $r$, which measures the accumulation of particles around the center of mass of the system, see Fig.~\ref{fig1}.a.  
Such DPT breaks the continuous time-translational symmetry of the original action, thus opening the door to its use as a resource to build continuous time crystals.

\begin{figure}
\includegraphics[width=9cm]{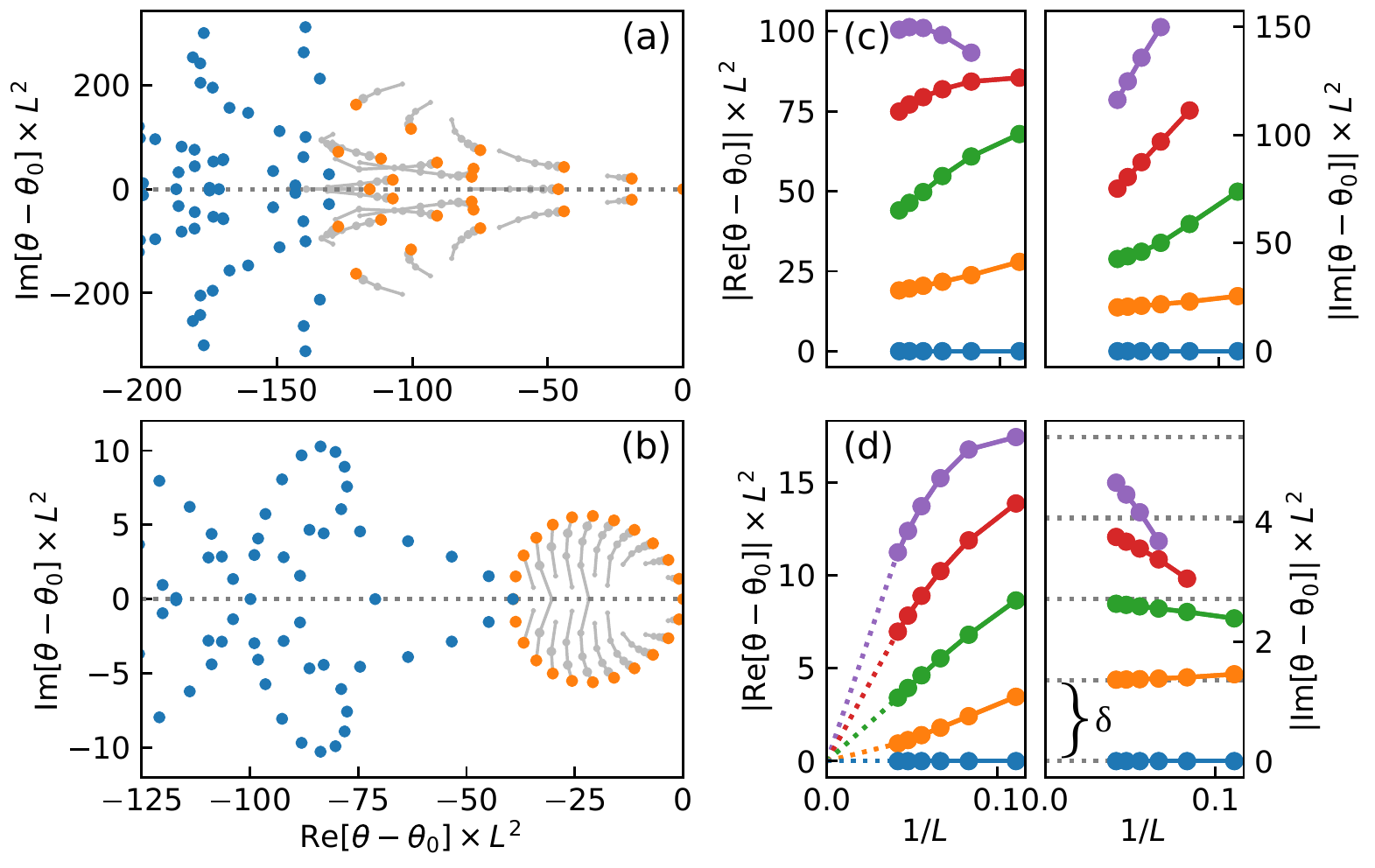}
\vspace{-0.6cm}
\caption{
    Diffusively-scaled spectrum of the tilted generator $\mathbb{W}^{\lambda}$ for $E=10$. (a) Homogeneous phase for $\lambda=-1$. (b) Condensate phase for $\lambda=-9$. Big colored points correspond to $L=24$, while small light gray points represent the leading eigenvalues for smaller lattice sizes ($L=9, 12, 15, 18, 21$), showing their evolution as $L$ increases. (c)-(d) Finite-size scaling analysis for the real and imaginary parts of the leading eigenvalues in the homogeneous (c) and condensate (d) phases. The real parts converge to zero as a power law of $1/L$ in the condensate phase, while the imaginary parts exhibit a clear band structure with constant frequency spacing $\delta$, proportional to the condensate velocity.}
\label{fig2}
\end{figure}

In this Letter we report three main results. Firstly we demonstrate that the rotating condensate corresponds to a time-crystal phase at the fluctuating level. We do this by exploring the spectral fingerprints of the DPT present in the WASEP. In particular, we show that the spectrum of the tilted generator describing current fluctuations in this model becomes asymptotically gapless for currents below a critical threshold. Here, a macroscopic fraction of eigenvalues shows a vanishing real part of the gap as the system size $L\to\infty$, while developing a band structure in the imaginary axis, see Fig.~\ref{fig2}, which is the hallmark of time crystals \cite{iemini18a}. Interestingly, these rare events can be made typical (i.e. a steady-state property) by virtue of Doob's transform \cite{doob57a, chetrite15a, chetrite15b, bertini15a, carollo18b,simon09a,jack10a,popkov10a}, which can be interpreted in terms of the original dynamics supplemented with a \emph{smart} driving field. The second main result consists in showing that this smart field acts as a \emph{packing field}, pushing particles that lag behind the condensate's center of mass while restraining those moving ahead. This amplifies naturally-occurring fluctuations of the packing parameter (see Fig.~\ref{fig1}.b), a nonlinear feedback mechanism (formally reminiscent of the Kuramoto synchronization transition \cite{kuramoto84a,kuramoto87a,pikovsky03a,acebron05a}) which eventually leads to a time-crystal phase. These observations lead us to the third main result, which distills the key properties of Doob's smart field to introduce the time-crystal lattice gas (tcLG). Numerical simulations and a local stability analysis of its hydrodynamics confirm that the tcLG exhibits a steady-state phase transition to a time crystalline phase with a matter wave which breaks continuous time-translation symmetry and displays rigidity, robust coherent periodic motion and long-range spatio-temporal order despite the stochasticity of the underlying dynamics. 

\textbf{\emph{Model.}}-- 
The WASEP belongs to a broad class of driven diffusive systems of fundamental interest \cite{de-masi89a,gartner87a,derrida07a}. Microscopically it consists of $N$ particles evolving in a $1d$ lattice of $L\ge N$ sites subject to periodic boundary conditions, so the total density is $\rho_0=N/L$. Each lattice site may be empty or occupied by one particle at most, so a microscopic configuration is given by $C=\{n_k\}_{k=1,\ldots,L}$ with $n_k=0,1$ the occupation number of the $k^{\text{th}}$ site and $N=\sum_{k=1}^L n_k$. Particles may hop randomly to empty neighboring sites along the $\pm x$-direction with rates $p_\pm = \frac{1}{2}\text{e}^{\pm E/L}$, with $E$ an external field which drives the system to a nonequilibrium steady state characterized by an average current $\la q\ra =\rho_0(1-\rho_0) E$ and a homogeneous density profile $\la n_k\ra = \rho_0~\forall k$. Configurations can be encoded as vectors in a Hilbert space \cite{schutz01a}, $\ket{C}=\bigotimes_{k=1}^L (n_k, 1-n_k)^T$, with $^T$ denoting transposition, and the system information at time $t$ is stored in a vector $\ket{P_t}=(P_t(C_1),P_t(C_2),...)^T=\sum_i P_t(C_i)\ket{C_i}$, with $P_t(C_i)$ representing the probability of configuration $C_i$. This probability vector is normalized, $\langle-|P_t\rangle=1$, with $\bra{-}=\sum_i\bra{C_i}$ and $\langle C_i| C_j \rangle=\delta_{ij}$. $\ket{P_t}$ evolves in time according to a master equation $\partial_t\ket{P_t}={\mathbb W} \ket{P_t}$, where ${\mathbb W}$ defines the Markov generator of the dynamics (see below). At the macroscopic level, driven lattice gases like WASEP are characterized by a density field $\rho(x,t)$ which obeys a hydrodynamic equation \cite{spohn12a}
\be
\partial_t \rho = -\partial_x \Big(-D(\rho)\partial_x \rho + \sigma(\rho) E \Big) \, ,
\label{langevin1}
\ee
with $D(\rho)$ and $\sigma(\rho)$ the diffusivity and mobility coefficients, which for WASEP are $D(\rho)=1/2$ and $\sigma (\rho)=\rho (1-\rho)$.

\textbf{\emph{Fluctuations.}}-- 
We consider now the statistics of an ensemble of trajectories conditioned to a given space- and time-integrated current $Q$ during a long time $t$. As in equilibrium statistical physics \cite{touchette09a}, this trajectory ensemble is fully characterized by a \emph{dynamical partition function} $Z_t(\lambda)=\sum_Q P_t(Q) \text{e}^{\lambda Q}$, where $P_t(Q)$ is the probability of trajectories of duration $t$ with total current $Q$, or equivalently by the associated \emph{dynamical free energy} $\theta(\lambda)=\lim_{t\to \infty} t^{-1} \ln Z_t(\lambda)$. 
The variable $\lambda$ is an intensive \emph{biasing field}, conjugated to the extensive current $Q$ in a way similar to the relation between temperature and energy in equilibrium systems \cite{bertini15a}. Negative (positive) values of $\lambda$ bias the statistics of $Q$ towards currents lower (larger) than the average stationary value, which corresponds to $\lambda=0$ \cite{garrahan09a}.
The statistics of the configurations associated with a rare event of parameter $\lambda$ are captured by a vector $\ket{P_t(\lambda)}$, which evolves in time according to a deformed master equation $\partial_t\ket{P_t(\lambda)}={\mathbb W}^{\lambda} \ket{P_t(\lambda)}$, with ${\mathbb W}^{\lambda}$ a \emph{tilted generator} which biases the original dynamics in order to favor large or low currents according to the sign of $\lambda$.
It can be shown \cite{touchette09a,hurtado14a,garrahan18a} that $\theta(\lambda)$ is the largest eigenvalue of ${\mathbb W}^{\lambda}$, as $Z_t(\lambda)=\bracket{-}{P_t(\lambda)}$. For WASEP \cite{lecomte07c,garrahan09a} 
\ben
{\mathbb W}^{\lambda}&=&\sum_{k=1}^{L}[\frac{1}{2}e^{\frac{\lambda+E}{L}} \hat{\sigma}_{k+1}^+\hat{\sigma}_k^- + \frac{1}{2}e^{-\frac{\lambda+E}{L}} \hat{\sigma}_{k}^+\hat{\sigma}_{k+1}^- \\ \nonumber
&-&\frac{1}{2}e^{\frac{E}{L}}{\hat n}_k ({\mathbb I}-{\hat n}_{k+1} )-\frac{1}{2}e^{-\frac{E}{L}}{\hat n}_{k+1} ({\mathbb I}-{\hat n}_k )]\, ,
\label{Wlambda}
\een
where $\hat{\sigma}_k^\pm$ are creation and annihilation operators acting on site $k\in[1,L]$, ${\mathbb I}$ is the identity matrix and ${\hat n}_k=\hat{\sigma}_k^+ \hat{\sigma}_k^-$ is the number operator. Note that the original Markov generator is just ${\mathbb W}\equiv {\mathbb W}^{\lambda=0}$, while ${\mathbb W}^{\lambda\ne 0}$ does not conserve probability (i.e. $\bra{-}{\mathbb W}^{\lambda\ne 0}\ne 0$). 

\textbf{\emph{Spectral analysis of the DPT.}}-- 
The WASEP has been shown to exhibit a DPT \cite{bodineau05a,perez-espigares13a,hurtado14a} to a time-translation symmetry-broken phase for $|E|>E_c\equiv \pi/\sqrt{\rho_0(1-\rho_0)}$ and $\lambda_c^-<\lambda<\lambda_c^+$, with $\lambda_c^\pm=\pm\sqrt{E^2-E^2_c}-E$, where $\theta(\lambda)$ develops a second-order singularity and a macroscopic jammed condensate emerges to hinder particle transport and thus aid low current fluctuations, see bottom inset in Fig.~\ref{fig1}.a. This DPT is well-captured by the packing order parameter $r(\lambda)$, the $\lambda$-ensemble average of $r_C\equiv |z_C|$, with $z_C\equiv N^{-1} \sum_{k=1}^N \text{e}^{i2\pi x_k(C)/L}= r_C \text{e}^{i\phi_C}$ and $x_k(C)$ the lattice position of particle $k$ in configuration $C$, see Fig.~\ref{fig1}.a. Note that $r_C=|z_C|$ and $\phi_C=\arg(z_C)$ are the well-known Kuramoto order parameters of synchronization \cite{kuramoto84a,kuramoto87a,pikovsky03a,acebron05a}, measuring in this case the particles' spatial coherence and the center-of-mass angular position, respectively, thus capturing the transition from the homogeneous to the density wave phase. The spectrum of ${\mathbb W}^{\lambda}$ codifies all the information on this DPT. In particular, let $\ket{R_i^\lambda}$ and $\bra{L_i^\lambda}$ be the $i^{\text{th}}$ ($i=0,1,\ldots,2^L-1$) right and left eigenvectors of ${\mathbb W}^{\lambda}$, respectively, so ${\mathbb W}^{\lambda}\ket{R_i^\lambda} = \theta_i(\lambda) \ket{R_i^\lambda}$ and $\bra{L_i^\lambda}{\mathbb W}^{\lambda} = \theta_i(\lambda) \bra{L_i^\lambda}$, with $\theta_i(\lambda)\in {\mathbb C}$ the associated eigenvalue ordered according to their real part (largest first), so that $\theta(\lambda)=\theta_0(\lambda)$. Fig. \ref{fig2}.a-b shows the spectrum of ${\mathbb W}^{\lambda}$ for $L=24$, $\rho_0=1/3$, $E=10$ and two values of the biasing field $\lambda$, one subcritical (Fig. \ref{fig2}.a) and another once the DPT has kicked in (Fig. \ref{fig2}.b). Clearly, the structure of the spectrum in the complex plane changes radically between the two phases. In particular, while the spectrum is gapped (in the sense that $\text{Re}[\theta_i-\theta_0]<0$ for $i>0$) for any $\lambda<\lambda_c^-$ or $\lambda>\lambda_c^+$ (Fig.~\ref{fig2}.c), the condensate phase ($\lambda_c^-<\lambda<\lambda_c^+$) is characterized by a vanishing gap in the real part of a macroscopic fraction of eigenvalues as $L\to\infty$, which decays as a power-law with $1/L$, see Fig.~\ref{fig2}.d. Moreover, the imaginary parts of the gap-closing eigenvalues exhibit a clear band structure with a constant frequency spacing $\delta$ which can be directly linked with the velocity $v$ of the moving condensate, $\delta=2\pi v/L$ (see dashed horizontal lines in Fig.~\ref{fig2}.d), all standard features of a time-crystal phase \cite{zakrzewski12a,moessner17a,richerme17a,yao18a,sacha18a}. Indeed, the emergence of a multiple (${\cal O}(L)$-fold) degeneracy as $L$ increases for $\lambda_c^-<\lambda<\lambda_c^+$ signals the appearance of different competing (symmetry-broken) states, related to the invariance of the condensate against integer translations along the lattice. This DPT at the fluctuating level has therefore the fingerprints of a time-crystal phase, thus enabling a path to engineer these novel phases of matter in driven diffusive systems.

\begin{figure}
\vspace{-0.4cm}
\includegraphics[width=8.5cm]{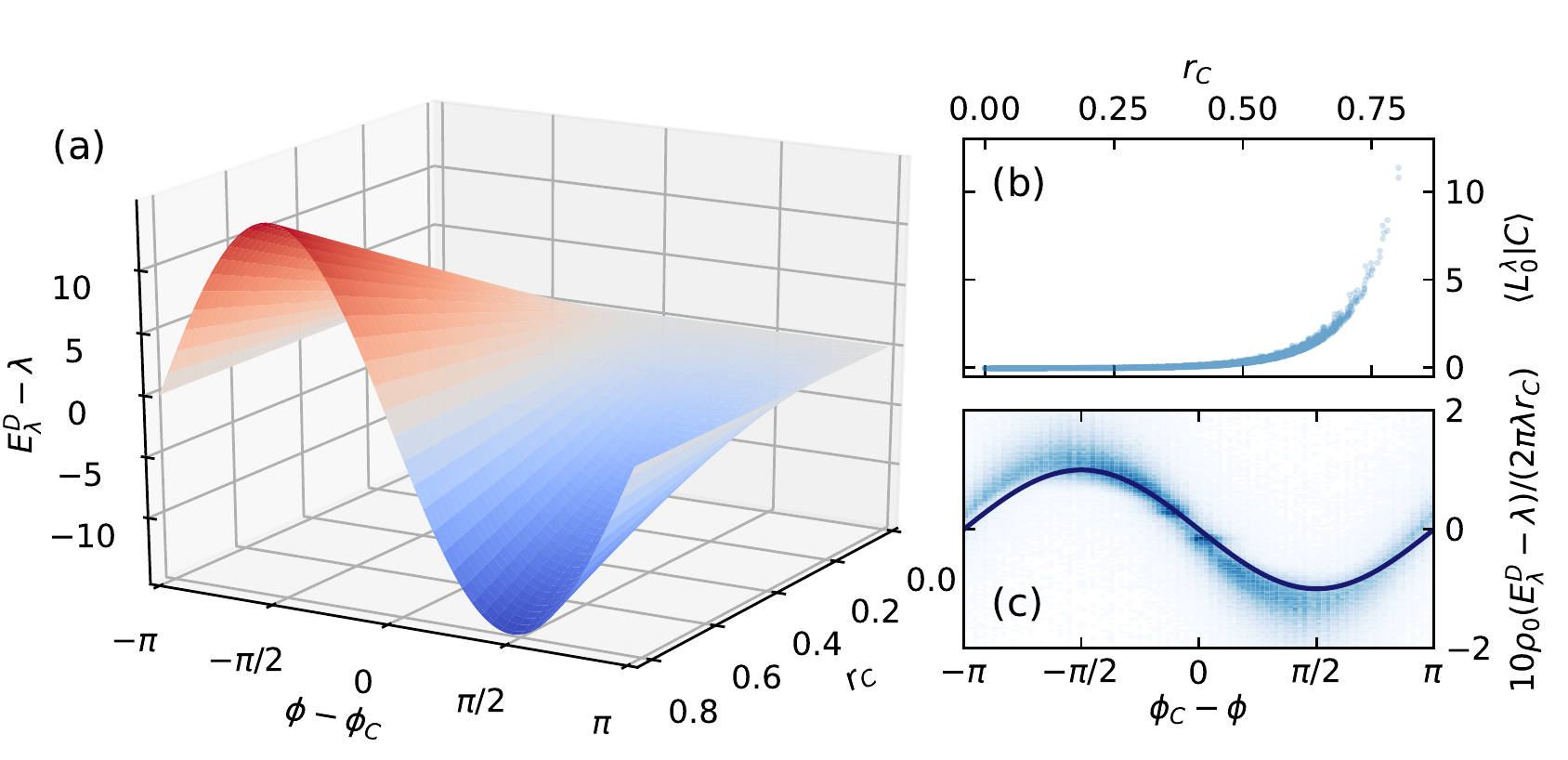}
\vspace{-0.2cm}
\caption{(a) Smart packing field for $\rho_0=1/3$ and $\lambda=-9$ as a function of packing order parameter $r_C$ and the angular distance to the center-of-mass position. (b) $\bracket{L_0^\lambda}{C}$ vs the packing order parameter $r_C$ for $L=24$, $\rho_0=1/3$, $E=10$, $\lambda=-9$ (condensate phase) and a large sample of microscopic configurations. (c) Angular dependence of the Doob's smart field with respect to the center-of-mass angular location for a large sample of microscopic configurations and the same parameters, together with the $\sin(\phi_k-\phi_C)$ prediction (line).}
\label{fig3}
\end{figure}

\textbf{\emph{Doob's smart field.}}-- 
We can now turn the condensate dynamical phase into a true time-crystal phase of matter by making typical the rare events for any $\lambda$, i.e. by transforming the non-stochastic generator ${\mathbb W}^{\lambda}$ into a physical generator ${\mathbb W}^{\lambda}_{\text{D}}$ via the Doob's transform ${\mathbb W}^{\lambda}_{\text{D}}\equiv {\mathbb L}_0{\mathbb W}^{\lambda}{\mathbb L}_0^{-1}-\theta_0(\lambda)$, with ${\mathbb L}_0$ a diagonal matrix with elements $({\mathbb L}_0)_{ii}= (\bra{L_0^{\lambda}})_i$ \cite{doob57a, chetrite15a, chetrite15b, bertini15a, carollo18b,simon09a,jack10a,popkov10a}.
${\mathbb W}^{\lambda}_{\text{D}}$ is now a probability-conserving stochastic matrix, $\bra{-}{\mathbb W}^{\lambda}_{\text{D}}=0$, with a spectrum simply related to that of ${\mathbb W}^{\lambda}$, i.e. $\theta_i^{\text{D}}(\lambda)=\theta_i(\lambda)-\theta_0(\lambda)$ with $\ket{R_{i,{\text{D}}}^\lambda}=\mathbb{L}_0\ket{R_{i}^\lambda}$ and $\bra{L_{i,{\text{D}}}^\lambda}=\bra{L_{i}^\lambda}\mathbb{L}_0^{-1}$, generating in the steady state the same trajectory statistics as ${\mathbb W}^{\lambda}$.
%To better understand the underlying physics, we now write Doob's dynamics in terms of the original WASEP dynamics supplemented by a \emph{smart} field $E^{\text{D}}_\lambda$, i.e. we define $({\mathbb W}^{\lambda}_{\text{D}})_{ij}=(\mathbb W)_{ij} \exp[q_{C_i C_j} (E^{\text{D}}_\lambda)_{ij}/L]$ with $({\mathbb W}^{\lambda}_{\text{D}})_{ij}=\bra{C_i}{\mathbb W}^{\lambda}_{\text{D}}\ket{C_j}$ and $q_{C_i C_j}=\pm 1$ the particle current involved in the transition $C_j\to C_i$.
To better understand the underlying physics, we now write Doob's dynamics in terms of the original WASEP dynamics supplemented by a \emph{smart} field $E^{\text{D}}_\lambda$, i.e. we define $({\mathbb W}^{\lambda}_{\text{D}})_{ij}=(\mathbb W)_{ij} \exp[\jumpdir_{C_i C_j} (E^{\text{D}}_\lambda)_{ij}/L]$ with $({\mathbb W}^{\lambda}_{\text{D}})_{ij}=\bra{C_i}{\mathbb W}^{\lambda}_{\text{D}}\ket{C_j}$ and $\jumpdir_{C_i C_j}=\pm 1$ the direction of the particle jump in the transition $C_j\to C_i$.
Together with the definition of ${\mathbb W}^{\lambda}_{\text{D}}$, this leads to
\be
(E^{\text{D}}_\lambda)_{ij} = \lambda + \jumpdir_{C_i C_j}L \ln \left( \frac{\bracket{L_0^\lambda}{C_i}}{\bracket{L_0^\lambda}{C_j}} \right) \, .
\label{EDoob}
\ee
$E^{\text{D}}_\lambda$ can be interpreted as the external field needed to make typical a rare event of bias field $\lambda$. In order to disentangle the nonlocal complexity of Doob's smart field, we scrutinize its dependence on the packing parameter $r_C$. In particular, Fig.~\ref{fig3}.b plots the projections $\bracket{L_0^\lambda}{C}$ vs the packing parameter $r_C$ for a large sample of microscopic configurations $C$, as obtained for $L=24$, $\rho_0=1/3$ and $\lambda=-9$ (condensate phase). Interestingly, this shows that $\bracket{L_0^\lambda}{C}\simeq f_{\lambda,L}(r_C)$ to a high degree of accuracy, with $f_{\lambda,L}(r)$ some unknown $\lambda$- and $L$-dependent function of the packing parameter. This means in particular that the Doob's smart field $(E^{\text{D}}_\lambda)_{ij}$ depends essentially on the packing parameter of configurations $C_i$ and $C_j$, a radical simplification. Moreover, as elementary transitions involve just a local particle jump, the resulting change on the packing parameter is perturbatively small for large enough $L$. In particular, if $C'_k$ is the configuration that results from $C$ after a particle jump at site $k\in[1,L]$, we have that $r_{C'_k}\simeq r_C + 2\pi \jumpdir_{C'_k C} (\rho_0 L^2)^{-1} \sin(\phi_C-\phi_k)$, with $\phi_k\equiv 2\pi k/L$. The Doob's smart field for this transition is then $(E^{\text{D}}_\lambda)_{C'_k,C}\simeq \lambda + 2\pi (\rho_0 L)^{-1} g_{\lambda,L}(r_C) \sin(\phi_C-\phi_k)$, with $g_{\lambda,L}(r)\equiv f'_{\lambda,L}(r)/f_{\lambda,L}(r)$, and we empirically find a linear dependence $g_{\lambda,L}(r)\approx -\lambda L r/10$ near the critical point $\lambda_c^+$. This is confirmed in Fig.~\ref{fig3}.c, where we plot $10\rho_0 [(E^{\text{D}}_\lambda)_{C'_k,C}-\lambda]/(2\pi \lambda r_C)$ obtained from Eq.~(\ref{EDoob}) for a large sample of connected configurations $C\to C'_k$ as a function of $\phi_C-\phi_k$. Similar effective potentials for atypical fluctuations have been found in other driven systems \cite{kaviani20a,popkov10a}. In this way, $(E^{\text{D}}_\lambda-\lambda)$ acts as a \emph{packing field} on a given configuration $C$, pushing particles that lag behind the center of mass while restraining those moving ahead, see Fig. \ref{fig3}.a, with an amplitude proportional to the packing parameter $r_C$ and $\lambda$. This nonlinear feedback mechanism, which competes with the diffusive tendency to flatten profiles and the pushing constant field, amplifies naturally-occurring fluctuations of the packing parameter, leading eventually to a time-crystal phase for $\lambda_c^-<\lambda<\lambda_c^+$. 

\begin{figure}[t]
\includegraphics[width=8.5cm]{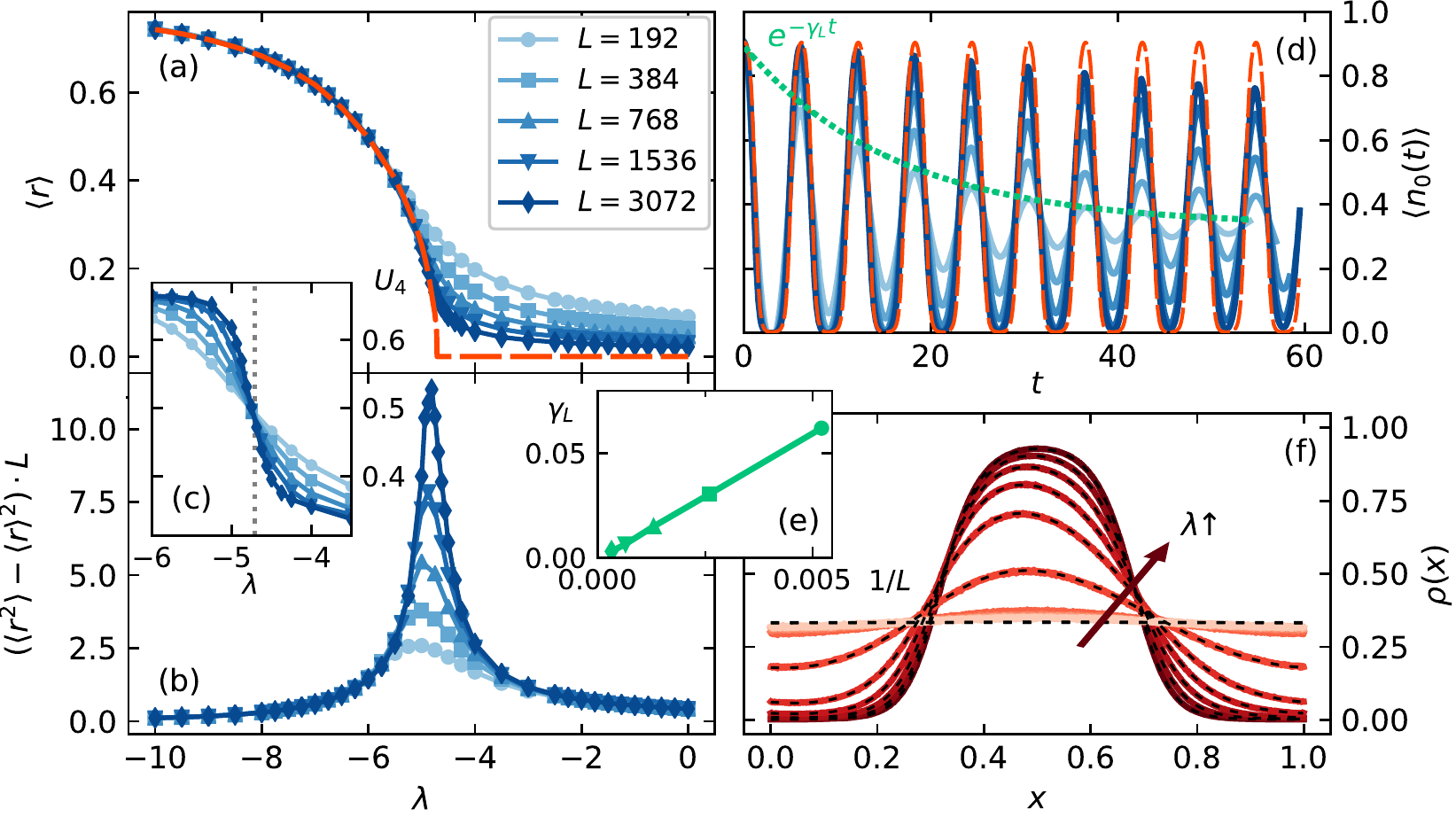}
\vspace{-0.25cm}
\caption{Numerics for the time-crystal lattice gas. Average packing order parameter (a), its fluctuations (b) and Binder's cumulant (c) measured for $\rho_0=1/3$, $E=10$ and different $L$. (d) Local density as a function of time and different $L$'s in the time-crystal phase ($\lambda=-9$). Note the persistent oscillations typical of time crystals. (e) Decay of the oscillations damping rate as $L\to\infty$, a clear sign of the rigidity of the time-crystal phase in the thermodynamic limit. (f) Average density profile of the condensate for $L=3072$ and varying $\lambda$. Dashed lines correspond to hydrodynamic predictions.}
\label{fig4}
\end{figure}

\textbf{\emph{Time-crystal lattice gas}.}-- Inspired by the results of the previous analysis, we now simplify the Doob's smart field to introduce the time-crystal lattice gas (tcLG). This is a variant of the $1d$ WASEP where a particle at site $k$ hops stochastically under a configuration-dependent packing field $E_\lambda(C;k)= E + \lambda + 2 \lambda r_C \sin(\phi_k-\phi_C)$, with $E$ being a constant external field and $\lambda$ now a control parameter. We note that this smart field can be also written as a Kuramoto-like long-range interaction term $E_\lambda(C;k)=E+\lambda+\frac{2\lambda}{N}\sum_{j\neq k} \sin(\phi_k-\phi_j)$, highlighting the link between the tcLG and the Kuramoto model of synchronization \cite{kuramoto84a, kuramoto87a, pikovsky03a, acebron05a}. However, we stress that this link is only formal, as Kuramoto model lacks any particle transport in real space. According to the discussion above, we expect this lattice gas to display a putative steady-state phase transition to a time-crystal phase with a rotating condensate at some critical $\lambda_c$ as $L \to \infty$ (due to the Perron-Frobenius theorem). To test this picture, we performed extensive Monte Carlo simulations and a finite-size scaling analysis of the tcLG at density $\rho_0=1/3$. The average packing parameter $\la r\ra$ increases steeply but continuously for $\lambda< \lambda_c=-\pi/(1-\rho_0)\approx -4.7$ see Fig.~\ref{fig4}.a, converging toward the macroscopic hydrodynamic prediction (see below) as $L\to\infty$. Moreover the associated susceptibility, as measured by the packing fluctuations $\la r^2\ra - \la r\ra^2$, exhibits a well-defined peak around $\lambda_c$ which sharpens as $L$ grows and is compatible with a divergence in the thermodynamic limit (Fig.~\ref{fig4}.b). The critical point location can be inferred from the crossing of the finite-size Binder cumulants $U_4(L)=1-\la r^4\ra/(3\la r^2\ra)$ for different $L$'s, see Fig.~\ref{fig4}.c, and agrees with the hydrodynamic value for $\lambda_c$. Interestingly, the average density at a given point exhibits persistent oscillations as a function of time with period $v^{-1}$ (in the diffusive timescale), see Fig.~\ref{fig4}.d, with $v$ the condensate velocity, a universal feature of time crystals \cite{yao18a, wilczek12a, shapere12a, zakrzewski12a, moessner17a, richerme17a, yao18a, sacha18a, bruno13a, nozieres13a, watanabe15a, khemani16a, keyserlingk16a, else16a, gambetta19a, yao17a, yao17a, zhang17a, choi17a, nakatsugawa17a, lazarides17a, gong18a, tucker18a, osullivan18a, lazarides19a, yao18b, gambetta19b, iemini18a, medenjak20a}, and converges toward the hydrodynamic (undamped) periodic prediction as $L\to \infty$. Indeed the finite-size damping rate of oscillations, $\gamma_L$, obtained from an exponential fit to the envelope of $\la n_0(t)\ra$, decays to zero in the thermodynamic limit (Fig.~\ref{fig4}.e), a clear signature of the rigidity of the long-range spatio-temporal order emerging in the time crystal phase of tcLG. We also measured the average density profile of the moving condensate, see Fig.~\ref{fig4}.f, which becomes highly nonlinear deep into the time-crystal phase. In the macroscopic limit, one can show using a local equilibrium approximation \cite{prados12a, hurtado13a, lasanta15a, lasanta16a, manacorda16a, gutierrez-ariza19a} that the tcLG is described by a hydrodynamic equation (\ref{langevin1}) with a $\rho$-dependent local field $E_\lambda(\rho;x)= E + \lambda + 2 \lambda r_\rho \sin(2\pi x-\phi_\rho)$, with $r_\rho=|z_\rho|$, $\phi_\rho=\arg(z_\rho)$, and $z_\rho=\rho_0^{-1}\int_0^1 dx \rho(x) \text{e}^{i2\pi x}$ the field-theoretic generalization of our complex order parameter. A local stability analysis then shows \cite{hurtado11a,hurtado14a,tizon-escamilla17b} that the homogeneous solution $\rho(x,t)=\rho_0$ becomes unstable at $\lambda_c=-2\pi\rho_0D(\rho_0)/\sigma(\rho_0)=-\pi/(1-\rho_0)$, where a ballistic condensate emerges. Hydrodynamic predictions are fully confirmed in simulations, see Fig. \ref{fig4}. Note that the tcLG hydrodynamics is similar to the continuous limit of the Kuramoto model \cite{acebron05a}, with the peculiarity that for tcLG the mobility $\sigma(\rho)$ is quadratic in $\rho$ (a reflection of microscopic  particle exclusion) while it is linear for Kuramoto.

\textbf{\emph{Conclusion.}}-- We provide here a new mechanism to engineer time-crystalline order in driven diffusive media by making typical rare trajectories that break time-translation symmetry,
%symmetry-breaking DPTs appearing at the fluctuating level in many-body systems, 
and physically based on the idea of a packing field which triggers a condensation instability. The modern experimental control of colloidal fluids trapped in quasi-$1d$ periodic structures, such as circular channels \cite{wei00a,lutz04a} or optical traps based e.g. on Bessel rings or optical vortices \cite{ladavac05a,roichman07a,roichman07b}, together with feedback-control force protocols to implement the nonlinear packing field $E_\lambda(C;k)$ using optical tweezers \cite{kumar18a,grier97a,ortiz14a,martinez15a,martinez17a,rodrigo18a}, may allow the engineering and direct observation of this time-crystal phase, opening the door to further experimental advances in this active field. Moreover, the ideas developed in this paper can be further exploited in $d>1$, where DPTs exhibit a much richer phenomenology \cite{tizon-escamilla17b,tizon-escamilla17a}, with different spatio-temporal symmetry-broken fluctuation phases separated by lines of $1^{\text{st}}$- and $2^{\text{nd}}$-order DPTs, competing density waves and coexistence. This may lead, via the Doob's transform pathway here described, to materials with a rich phase diagram composed of multiple spacetime-crystalline phases.

\begin{acknowledgments}
The authors thank F. Gambetta and Ra\'ul A. Rica for insightful discussions. The research leading to these results has received funding from the Spanish \emph{Ministerio de Econom\'{\i}a y Competitividad} project FIS2017-84256-P, EPSRC Grant No.~EP/N03404X/1 and from the European Regional Development Fund, Junta de Andalucía-Consejería de Economía y Conocimiento, Ref. A-FQM-175-UGR18. R.H.G. thanks funding from the Spanish \emph{Ministerio de Ciencia, Innovaci\'on y Universidades} fellowship FPU17/02191. C.P.E. acknowledges funding from the European Union's Horizon 2020 research and innovation programme under the Marie Sklodowska-Curie Cofund Programme Athenea3I Grant Agreement No.~754446. F.C. acknowledges support through a Teach@Tübingen Fellowship. We are also grateful for the computational resources and assistance provided by PROTEUS, the supercomputing center of the Institute Carlos I for Theoretical and Computational Physics at the University of Granada, Spain. 

C.P.E. and P.I.H. contributed equally to this work.
\end{acknowledgments}

\bibliography{referencias-BibDesk-OK}{}

\end{document}